\documentstyle[12pt]{article}
\input epsf
\begin{document}
\title{A new determination of polarized parton densities in the nucleon}
\author{ Jan
Bartelski\\ Institute of Theoretical Physics, Warsaw University,\\
Ho$\dot{z}$a 69, 00-681 Warsaw, Poland. \\ \\ \and Stanis\l aw
Tatur
\\ Nicolaus Copernicus Astronomical Center,\\ Polish Academy of
Sciences,\\ Bartycka 18, 00-716 Warsaw, Poland. \\ }
\date{}
\maketitle
\vspace{1cm}
\begin{abstract}
\noindent In order to determine polarized parton distributions we
have made a new NLO QCD fit using all experimental data on spin
asymmetries measured in the deep inelastic scattering on different
nucleon targets. The functional form of such densities is based on
MRST2001 results for unpolarized ones. We get for polarization of
quarks (at  $ Q^{2}=1\, {\rm GeV^{2}}$): $\Delta u = 0.87 , \Delta
d= -0.38, \Delta s= -0.04$. The total quark polarization is rather
big and we obtain: $\Delta\Sigma=0.44$. As a result of our fit we
get $a_{3} \cong g_{A}=1.24$, the value which is close to
experimental number. With negligible $\Delta s$  and  rather big
$\Delta\Sigma$ (comparable to $a_{8}$) the results of our new fit
are quite different in character from other fits.
\end{abstract}

\newpage

The experimental data on deep inelastic scattering of polarized
leptons on polarized nucleons was collected for many years in
experiments made in SLAC \cite{slac}, CERN  \cite{cern} and DESY
\cite{hermes}. The results were analyzed by many groups and next
to leading order (NLO) QCD polarized parton distributions were
determined \cite{fit,BT,BTpr, BTapp}. In the present paper we do
not consider any new experimental data. All the existing
experimental data on polarized deep inelastic scattering were
already included in our latest fits \cite{BTpr,BTapp}. However our
method of determination of spin densities depends very strongly on
the parton distributions for unpolarized case.  We assume that the
asymptotic behaviour of our polarized parton distributions is
determined (up to the condition that the corresponding parton
densities are integrable) by the fit to unpolarized data.

Recently a new determination of unpolarized parton densities
performed by Martin, Roberts, Stirling and Thorne (MRST2001)
\cite{MRSTnewn} was published. These distributions have
substantially modified (compared to older MRST98 \cite{MRSTnew}
fit) small $x$ behaviour of valence $u$ quark and gluon, as well
as gluon density is not positive for all values of $x$ variable.
We want to update our fit and to check how the new functional form
of parton distributions influences it. We will  follow the method
presented in \cite{BT, BTpr, BTapp} where we use $\overline{MS}$
renormalization scheme in QCD. Despite of the fact that we only
modify the functional dependence of the fitted (at $ Q^{2}=1\,
{\rm GeV^{2}}$) parton densities the obtained polarizations of
quarks (i.e., $\Delta u, \Delta d$ and $\Delta s$) and gluons
($\Delta G$) are significantly changed and the value of total
quark polarization $\Delta\Sigma$ is increased.  We will use data
for spin asymmetry at given $x$ and different $Q^2$ (431
experimental points).

Experiments on unpolarized targets provide information on the quark
 densities $q(x,Q^2)$ and $G(x,Q^2)$ inside the nucleon.
These densities can be expressed in term of $q^{\pm}(x,Q^2)$ and
$G^{\pm}(x,Q^2)$, i.e. densities of quarks and gluons with
helicity along or opposite to the helicity of the parent nucleon:
\begin{equation}
q = q^{+}+q^{-},\hspace*{1.5cm} G = G^{+}+G^{-}.
\end{equation}

The polarized parton densities, i.e. the differences of $q^{+}$,
$q^{-}$ and $G^{+}$, $G^{-}$ are given by:
\begin{equation}
\Delta q = q^{+}-q^{-},\hspace*{1.5cm} \Delta G = G^{+}-G^{-}.
\end{equation}

We will try to determine $\Delta q(x,Q^2)$ and $\Delta G (x,Q^2)$,
keeping in mind relations between eqs. (1) and (2).

The formulas for unpolarized quark and gluon distributions
determined (at $ Q^{2}=1\,{\rm GeV^{2}}$) in the fit performed by
Martin, Roberts, Stirling and Thorne \cite{MRSTnewn} (they use
$\Lambda^{n_{f}=4}_{\overline{MS}}=0.323$ $\mbox{GeV}$ and
$\alpha_s(M^2_Z)=0.119$) are:

\begin{eqnarray}
\noindent u_{v}(x)&=&0.158 x^{-0.75}(1-x)^{3.33}(1+5.61\sqrt {x}+55.49 x), \nonumber \\
\noindent d_{v}(x)&=&0.040 x^{-0.73}(1-x)^{3.88}(1+52.73\sqrt{x}+30.65 x), \nonumber \\
\noindent 2\bar{u} (x)&=&0.4 M(x)-\delta (x), \\ \noindent 2\bar{d} (x)&=&0.4 M(x)+\delta (x), \nonumber \\
\noindent  2\bar{s} (x)&=&0.2 M(x), \nonumber \\
\noindent G (x)&=&1.90x^{-0.91}(1-x)^{3.70}(1+1.26 \sqrt{x}-1.43 x) -0.21x^{-1.33}(1-x)^{10}. \nonumber
\end{eqnarray}
\noindent where:
\begin{eqnarray}
\noindent M(x)&=&0.222 x^{-1.26}(1-x)^{7.10}(1+3.42\sqrt{x}+10.3 x), \nonumber \\
\noindent  \delta (x)&=&1.195 x^{0.24}(1-x)^{9.10}(1+14.05 x-45.52 x^{2}).
\end{eqnarray}

We will split $q$ and $G$, as was already discussed in
\cite{BT,BTpr,BTapp},  into two parts in such a manner that the
quark distributions $q^{\pm}(x,Q^2)$  remain positive. Our
polarised densities  for valence quarks, sea antiquarks (the same
distribution we take for sea quarks) and gluons (at $
Q^{2}=1\,{\rm GeV^{2}}$) are parametrised as follows:
\begin{eqnarray}
\noindent \Delta u_{v}(x)&=&x^{-0.75}(1-x)^{3.33}(a_{1}+a_{2}\sqrt {x}+a_{4}x), \nonumber \\
\noindent \Delta d_{v}(x)&=&x^{-0.73}(1-x)^{3.88}(b_{1}+b_{2}\sqrt{x}+b_{3}x), \nonumber \\
\noindent 2\Delta \bar{u} (x)&=&0.4\Delta M(x)-\Delta \delta (x), \\ \noindent 2\Delta \bar{d} (x)&=&0.4\Delta M(x)+\Delta \delta (x), \nonumber \\
\noindent  2\Delta \bar{s} (x)&=&0.2\Delta M_s(x), \nonumber \\
\noindent \Delta G (x)&=&x^{-0.91}(1-x)^{3.70}(d_1+d_2 \sqrt{x}+d_3 x) +x^{-0.83}(1-x)^{10} d_4. \nonumber
\end{eqnarray}

\noindent where:
\begin{eqnarray}
\noindent \Delta
M(x)&=&x^{-0.76}(1-x)^{7.10}(c_{1}+c_{2}\sqrt{x}), \nonumber \\
\noindent \Delta M_s&=& x^{-0.76}(1-x)^{7.10}(c_{1s}+c_{2s}\sqrt{x}),  \\
\noindent \Delta \delta (x)&=&x^{0.24}(1-x)^{9.10}c_{3}(1+14.05x-45.52x^{2}). \nonumber
\end{eqnarray}
In order to have finite polarization for partons we use less
divergent distributions at $x \rightarrow 0$ for sea quarks and gluons (in the second term).
We also define:
\begin{eqnarray}
\noindent \Delta u&=&\Delta u_{v}+2 \Delta \bar{u} , \nonumber \\
\noindent \Delta d&=&\Delta d_{v}+2 \Delta \bar{d} ,  \\
\noindent \Delta s&=&2 \Delta \bar{s}, \nonumber
\end{eqnarray}
and
\begin{eqnarray}
\noindent \Delta \Sigma&=&\Delta u+ \Delta d +\Delta s , \nonumber \\
\noindent a_{8}&=&\Delta u+ \Delta d -2 \Delta s ,  \\
\noindent a_{3}&=&\Delta u- \Delta d=g_{A} . \nonumber
\end{eqnarray}

In order to determine the unknown parameters in the expressions
for  polarized quark and gluon distributions (eqs.(5,6)) we
calculate the spin asymmetries (starting from initial value $Q^2$
= 1 $\mbox{GeV}^2$) for measured values of $Q^2$ and make a fit to
the experimental data on spin asymmetries for proton, neutron and
deuteron targets. The spin asymmetry $A_1(x,Q^2)$ can be expressed
via the polarized structure function $g_1(x,Q^2)$ as:
\begin{equation}
A_1(x,Q^2)\cong \frac{(1+\gamma^{2}) g_{1}(x,Q^2)}{ F_1(x,Q^2)}=
\frac{ g_{1}(x,Q^2)}{ F_2(x,Q^2)}[2x(1+ R(x,Q^2))],
\end{equation}
\noindent where  $R = [F_2(1+\gamma^{2})-2xF_1]/ 2xF_1$, whereas
$\gamma =2Mx/Q$ ($M$ stands for proton mass). We will take the new
determined value of $R$ from the \cite{whitn}.  In calculating
$g_{1}(x,Q^2)$ and $F_{2}(x,Q^2)$ in the next to leading order we
use procedure described in \cite{BT, BTpr}.  Having calculated the
asymmetries according to eq.(9) for the  value of $Q^2$ obtained
in experiments we can make a fit to asymmetries on proton, neutron
and deuteron targets. The value of $a_3$ is not constrained in
such fit. We will also do not fix $a_{8}$  but we put a constraint
on its value. Simply we will add it as an extra experimental point
(from hyperon decays one has $a_8 = 0.58 \pm 0.1$, where we
enhance (to 3$\sigma$) an error).

Not all parameters are important in our fits.  We will  assume
that $c_{1s}=c_{1}$ (i.e. the most singular terms for strange and
non-strange sea contributions are equal). Such assumption
practically does not change the value of $\chi^{2}$ but improves
$\chi^{2}/N_{DF}$. In this case we get the following values of
parameters (at $Q^2 = 1\, \mbox{GeV}^2$) from the fit to all
existing data for  for spin asymmetries:
\begin{equation}
\begin{array}{llll}
a_{1}=\hspace*{0.45cm} 0.16 ,&a_{2}=\hspace*{0.15cm}-2.45,&a_{4}=\hspace*{0.45cm} 9.89,&\\
b_{1}=\hspace*{0.15cm} -0.04,&b_{2}=\hspace*{0.15cm}-1.03,&b_{3}=\hspace*{0.45cm} 0.65,&\\
c_{1}=\hspace*{0.15cm}-0.05,&c_{2}=\hspace*{0.45cm} 2.28,&c_{3}=\hspace*{0.45cm}1.20,&\\
c_{1s}\hspace*{0.0cm}=\hspace*{0.30cm}c_{1},&c_{2s}=\hspace*{0.0cm} -0.33,&\\
d_{1}=\hspace*{0.45cm} 8.67,&d_{2}=\hspace*{0.45cm}2.20,&d_{3}=\hspace*{0.15cm} -39.8,&d_{4}=-13.6.
\end{array}
\end{equation}

The quark and gluon distributions obtained from our fit  lead for
$ Q^{2}=1\,{\rm GeV^{2}}$ to the following integrated (over $x$)
quantities: $\Delta u = 0.87, \Delta d= -0.38, \Delta s= -0.04,
\Delta u_v = 0.77,   \Delta d_v = -0.67, 2\Delta \bar{u} = 0.10,
2\Delta \bar{d} = 0.29.$

We have positively polarized sea for up and down quarks and small
negatively  polarized sea for strange quarks. One can see
substantial breaking of $SU(2)$ symmetry in a sea. We got the
value of $a_3 \cong g_A=1.24$, the number which is very close to
experimental figure ($1.267 \pm 0.003$ \cite{pdg}).

As is seen from eqs.(5) and (6) the small $x$ behaviour of
expressions for valence  $u, d$ quarks and sea contribution
$\Delta M$ have very similar  behaviour at small values of $x$. So
one cannot expect that the splitting into valence and sea
contributions will be well determined in the fit. The integrated
values of quantities that are well determined i.e., $\Delta u =
0.87, \Delta d= -0.38,  \Delta s= -0.04, \Delta\Sigma=0.44$ and
$g_{A}=1.24$ can be
 compared with the values from previous fit  (called  $A'_{2}$
 in \cite{BTpr}) $\Delta u = 0.77, \Delta d= -0.57,
 \Delta s= -0.19, \Delta\Sigma=0.01$ and $g_{A}=1.34$.
 More singular behaviour of $\Delta u$ at small $x$ in the present
 fit does not only directly influence the value of $\Delta u$. We
 have some increase in $\Delta u$ and relatively big
 increase in $\Delta d$ and $\Delta s$ resulting in the higher
 value of $\Delta \Sigma$.
 It seems that improvement of the MRST2001 \cite{MRSTnewn}
 in comparison to MRST98 \cite{MRSTnew} changes significantly
 character of obtained results.
On the other hand the value of $\chi^2=376.7$ for the present fit
is higher then in our previous one \cite{BTpr}, despite of the
fact that we have used the same experimental data.

The  results in the "measured" region (i.e., for $0.003\le x \le
1$) are:  $\Delta u = 0.76, \Delta d= -0.32,  \Delta s= -0.03,
\Delta\Sigma=0.41$ and $g_{A}=1.08$ which should be  compared with
our previous results \cite{BTpr} of fit  $A'_{2}$  $\Delta u =
0.80 , \Delta d= -0.43,  \Delta s= -0.11, \Delta\Sigma=0.26$ and
$g_{A}=1.23$. In present fit we have the value of $\Delta G=25.7$
integrated in the whole region of $x$ and $\Delta G=-1.8$ in the
"measured" region of $x$, which have to be compared with results
of a $A'_{2}$ fit: $\Delta G=-0.19$ in the whole region  and the
same value  in the region: $0.003\le x \le 1$.

As we already mentioned before the assumption that $\Delta G$
should be integrable was used by us in basic fit and  the most
singular part of $G$ was removed from $\Delta G$ parametrization.
We have not assumed positivity for gluon contribution because
already in unpolarized case $G(x)$ was negative for small $x$
values. When we take for $\Delta G(x)$ the same functional form as
for $G(x)$ leaving most singular term for small $x$ (not
integrable) we get fit with smaller $\chi^2$, namely 372.8 (there
are some changes in parameters but the character of the fit does
not change). We get:  $\Delta u = 0.87, \Delta d= -0.38,
 \Delta s= -0.04, \Delta\Sigma=0.45$ and $g_{A}=1.24$. The value
 of gluon polarization  ($\Delta G$) is $-0.79$ in the measured
 region and of course infinite
in the whole region. We have made also a fit where we have not
assumed any positivity conditions for parton densities. It means
that functional form for such densities was used with arbitrary
(unconstrained) parameters in eqs.(5) and (6). We get, as
expected, much smaller $\chi^2$ value, namely 360.1 (smaller than
in our basic fit). The integration in  the whole $x$ region gives:
$\Delta u = 0.84, \Delta d= -0.38,  \Delta s= -0.04,
\Delta\Sigma=0.41$ and $g_{A}=1.22$. $\Delta G=-12.2$. When we
neglect the second term in gluon distribution (i.e., when we put
$d_4=0$) one gets $\chi^2=384.2$ and: $\Delta u = 0.85, \Delta d=
-0.40, \Delta s= -0.06, \Delta\Sigma=0.39$, $\Delta G=-14$.
 If we put  (for $ Q^{2}=1\,{\rm GeV^{2}}$) gluon polarization
 arbitrarily equal to zero (i.e. $d_{1}=d_{2}=d_{3}=d_{4}=0$)
 we get substantial increase in $\chi^2=393.2$,
whereas: $\Delta u = 0.84, \Delta d= -0.42,  \Delta s= -0.07,
\Delta\Sigma=0.36$ and $g_{A}=1.26$.

\begin{figure}
\noindent \hspace{-0.5cm} \epsfxsize =400pt\epsfbox{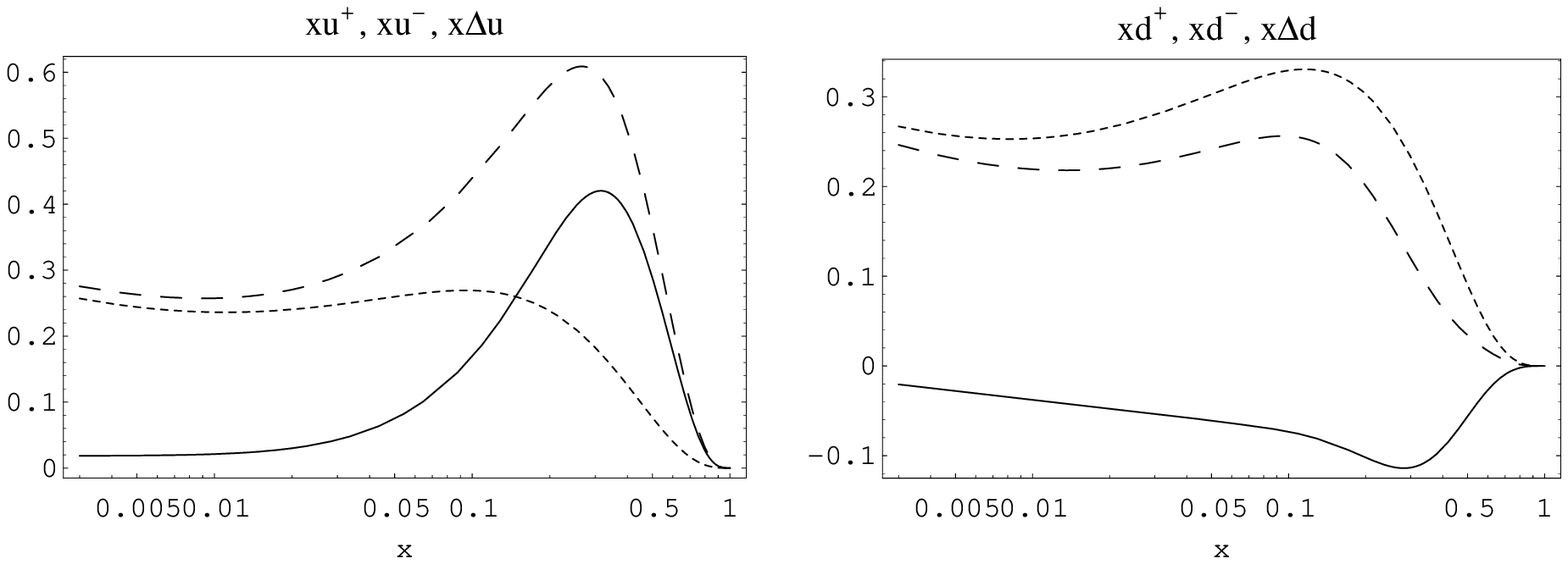}
\vspace{0.5cm}

\noindent \hspace{-0.5cm} \epsfxsize =400pt\epsfbox{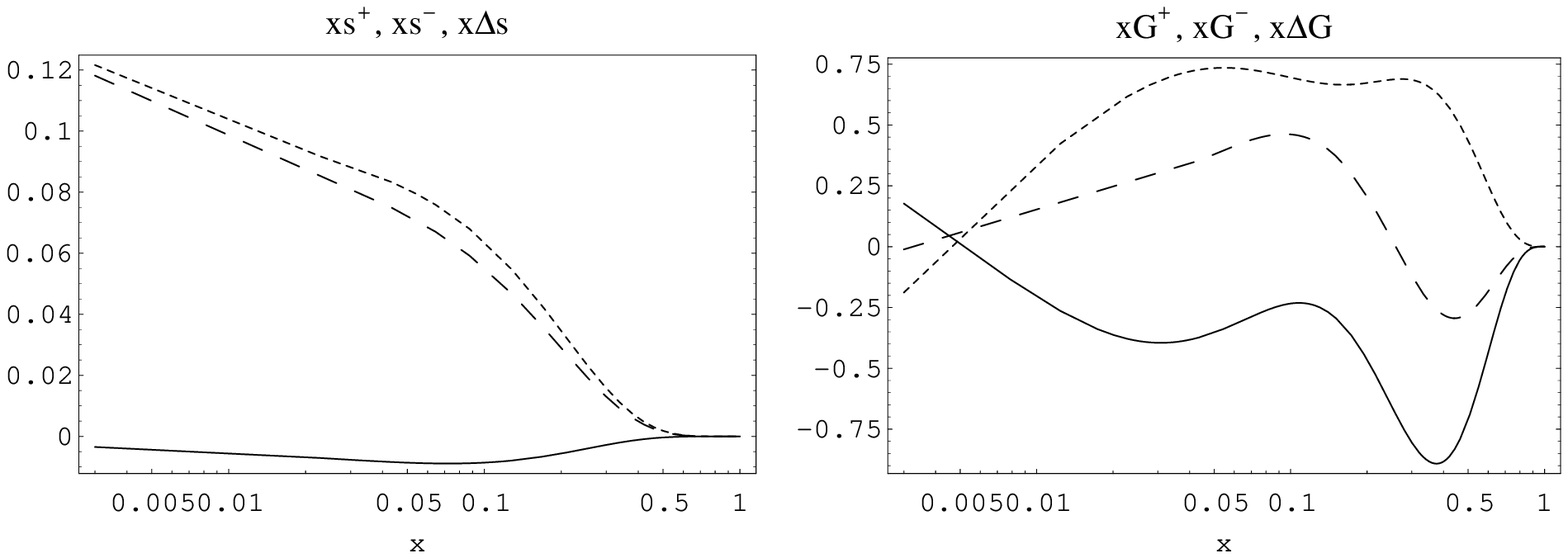}
\caption{\em{The quark (for different flavours) and gluon
densities versus $x$ obtained from our fit. The solid lines
represent parton polarization, whereas dashed (dotted) lines
correspond to $+$ ($-$) helicity component of parton
distribution.}}
\end{figure}

\begin{figure}
\noindent \hspace{-0.5cm} \epsfxsize =400pt\epsfbox{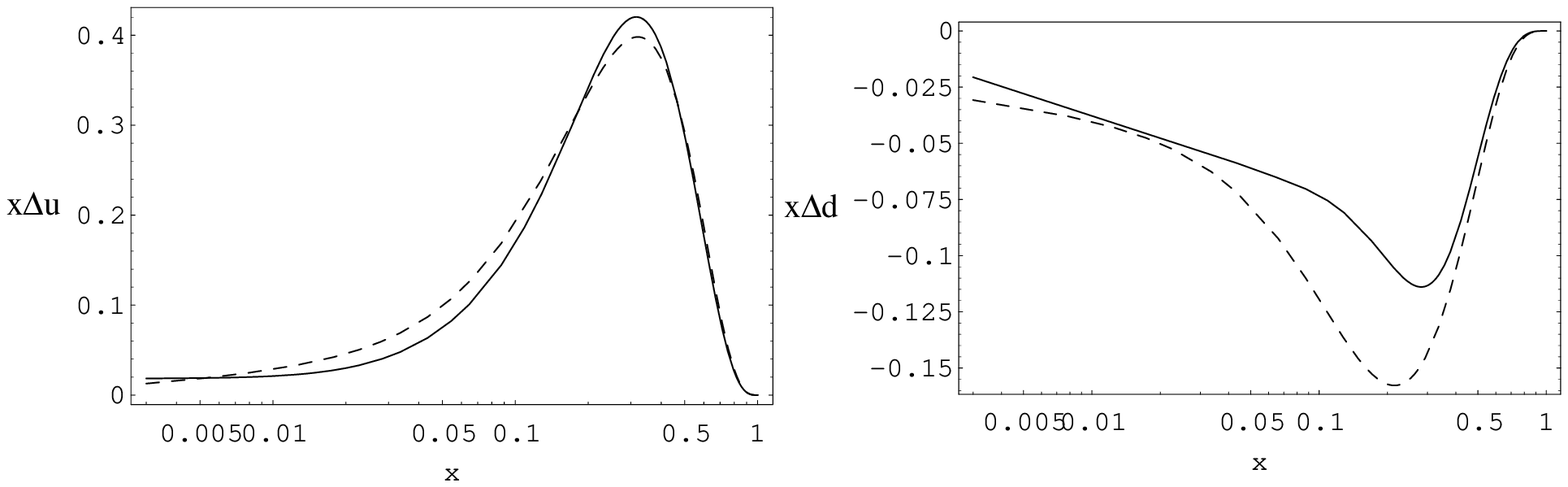}
\vspace{0.5cm}

\noindent \hspace{-0.5cm} \epsfxsize =400pt\epsfbox{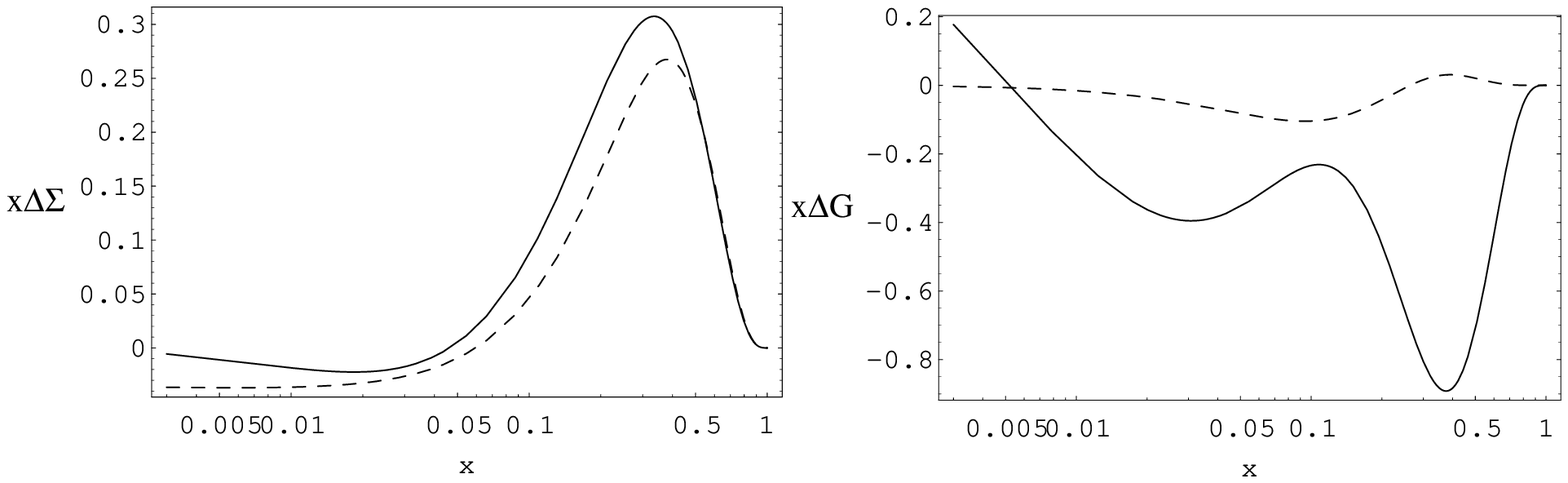}
\caption{\em{The parton densities versus $x$ obtained from our fit
(solid lines) compared with the results from our previous fit
\cite{BTpr}  (dashed lines).}}
\end{figure}

\begin{figure}
\noindent \hspace{-0.5cm} \epsfxsize =400pt\epsfbox{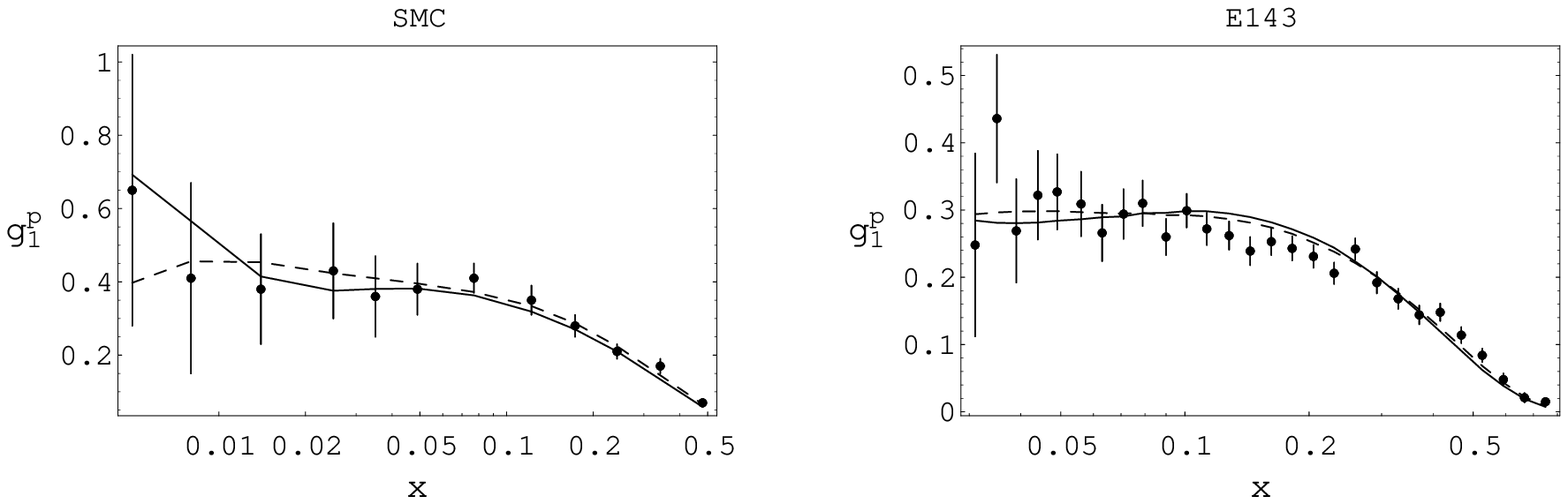}
\vspace{0.5cm}

\noindent \hspace{-0.5cm} \epsfxsize =400pt\epsfbox{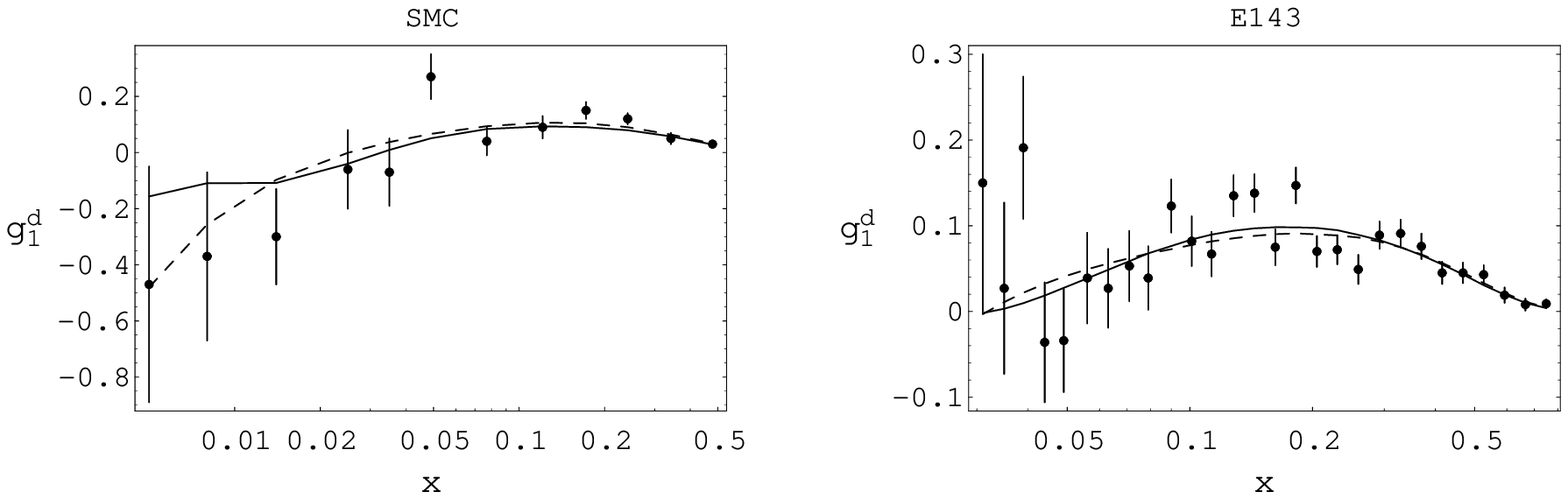}
\vspace{0.5cm}

 \noindent \hspace{-0.5cm} \epsfxsize
=400pt\epsfbox{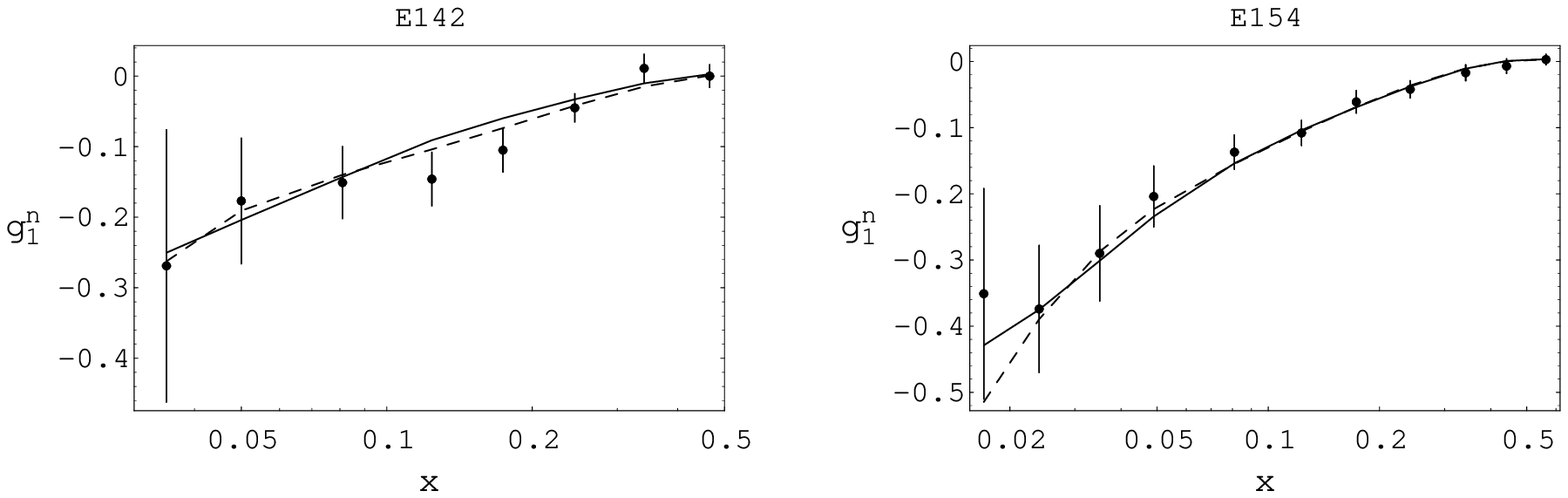} \caption{\em{The comparison of our
predictions for $g_1^N(x,Q^2)$ versus $x$ with the experimental
data from different experiments. Solid curves are calculated using
the distributions from our fit, the dashed ones are calculated
using the parameters of fit $A_{2}^{'}$ from \cite{BTpr}.}}
\end{figure}

It seems that the values of $\Delta u, \Delta d,  \Delta s,
\Delta\Sigma$ and $g_{A}$ do not change  much when we use
different assumptions about gluon
 contribution. We get in our fits small value of $\Delta  s$
 and $\Delta\Sigma$ comparable with $a_{8}$ (such conclusions
 remind Ellis-Jaffe
 results \cite{ej}). This is what differs our present results
 (coming from updated MRST2001 \cite{MRSTnewn} fit) from our
 previous fits \cite{BT,BTpr}.

Our fitting procedure also determines $+$ and $-$ helicity
components of parton densities. These components and the total
polarized distributions for quarks of different flavours and
gluons (calculated at $ Q^{2}=1\, {\rm GeV^{2}}$) are presented in
Fig.1.

In Fig.2 the values of functions of $\Delta u(x), \Delta d(x),
\Delta \Sigma(x)$ and $\Delta G(x)$ obtained in the present fit
are compared with the corresponding ones from our previous fit
called $A'_{2}$ \cite{BTpr}. One can see significant changes
(except maybe $\Delta u$) in these distributions.

In Fig.3 we present the comparison of  structure functions
$g_{1}^{p}$, $g_{1}^{d}$ and $g_{1}^{n}$ (calculated at the values
of $Q^2$ corresponding to data) with the corresponding
experimental points. For comparison, the curves for polarized
structure functions obtained in \cite{BTpr} from fit $A'_{2}$ are
also presented. The increases of $g_1$ (however within
experimental errors) are seen only for very small $x$ values  for
data from SMC experiments (on proton and deutrium targets) and for
E154 data on neutron target.

We have made fits to precise data on spin asymmetries on proton,
neutron and deuteron targets. Our model for polarized parton
distributions is
 based on new MRST2001 fit \cite{MRSTnewn} to unpolarized data.
  Because of different functional form of fitted parton densites (e.g. more
singular behaviour at $x=0$ for $u$ quark) the new fit is
different in character from the other fits \cite{fit,BT,BTpr}. We
have got relatively small value of $\Delta s$ and relatively high
value of $\Delta \Sigma$ (comparable with $a_{8})$ with $g_{A}$
(unconstrained) very close to experimental value. Gluon
contribution comes out relatively high and is (as usual) not very
reliable. It seems that in our model the problem of so called
"spin crisis" practically disappears.

\end{document}